\documentclass[11pt,notoc]{JHEP3}
   
\usepackage{epsfig}

\def\al{\alpha}

\def\de{\delta}

\def\la{\lambda}

\def\si{\sigma}

\def\Om{\Omega}

\newcommand{\ben}{\begin{equation}}
\newcommand{\een}{\end{equation}}
\newcommand{\bea}{\begin{eqnarray}}
\newcommand{\eea}{\end{eqnarray}}
\newcommand{\bit}{\begin{itemize}}
\newcommand{\eit}{\end{itemize}}

\def\pa{\partial}

\newcommand{\nlsm}{NL$\sigma$M}
\newcommand{\<}{\langle}
\renewcommand{\>}{\rangle}

\newcommand{\M}{\ensuremath{{\cal M}}}

\newcommand{\V}{\ensuremath{{\cal V}}}

\newcommand{\del}{\ensuremath{\partial}}

\newcommand{\half}{\ensuremath{\frac{1}{2}}}

\newcommand{\Tr}{\ensuremath{{\rm Tr}}}

\def\ZM{{\mathbb Z}}

\newcommand{\ba}{\begin{eqnarray}}
\newcommand{\ea}{\end{eqnarray}}

\title{Scaling in a SU(2)/$\ZM_3$ model of cosmic superstring networks}

\author{Mark Hindmarsh\\
Department of Physics and Astronomy\\
University of Sussex\\
Brighton BN1 9QH\\
U.K.}

\author{P.M. Saffin\\
School of Physics and Astronomy\\
University of Nottingham\\
University Park\\
Nottingham NG7 2RD\\
U.K.}

\abstract{
Motivated by recent developments in superstring theory in the cosmological context, 
we examine a field theory which contains string networks with 3-way junctions. 
We perform numerical simulations of this model, identify
the length scales of the network that forms, and provide evidence that the length scales tend towards a scaling regime, growing in proportion to time. We infer that the presence of junctions does not 
in itself cause a superstring network to dominate the energy density of the early Universe.
}


\begin{document}

\section{Introduction}
\label{sec:introduction}

Type IIB string theory, after compactification to 3+1 dimensions, has a 
spectrum of 1-dimensional objects  
in the form of solitonic D-strings and fundamental or F-strings which are of great 
current interest in cosmology 
\cite{Copeland:2003bj,Polchinski:2004ia}.
Such objects offer a much richer form of dynamics
than the more commonly studied Nielson-Olesen string of the Abelian Higgs model.
Individually, the F and D strings are $\half$-BPS objects, but they each break a different half
of the supersymmetry. When combined in a bound state one may think that this would further
break the supersymmetry, but in fact the bound states are still $\half$-BPS and give the
familiar BPS tension for bound states, 
\ba
\mu_{(p,q)} = \mu_F\sqrt{p^2 + q^2/g_s^2}.
\label{e:PQtension}
\ea
Where we have $p$ F-strings and $q$ D-strings, $\mu_F$ is the 
effective fundamental string tension after compactification and $g_s$ is the string coupling.
Given that stable bound states of vortices are present we should expect that junctions
will also be in the spectrum, where two different types of string come together at a point
and form a bound state leading away from that point. Such junctions not only exist,
but are $\frac{1}{4}$-BPS states \cite{Schwarz:1996bh,Dasgupta:1997pu}. 
In fact, it is possible to construct whole webs of
these junctions and still preserve supersymmetry \cite{Sen:1997xi}.
This is not the whole story as far as stability is concerned, for example we know that
strings can break by ending on a D-brane. In \cite{Copeland:2003bj} the authors give a
comprehensive account of the mechanisms which can lead to cosmic superstring breaking,
concluding that it is possible for stable cosmic-superstrings to exist.
Knowing that such strings can exist is not the same as saying they do: 
a mechanism is required to generate them. An attractive scenario is provided 
by the annihilation of a D3-$\overline\textrm{D3}$ pair terminating a period of brane 
inflation, which results in the formation of a networks of $(p,q)$-strings 
\cite{Sarangi:2002yt,Jones:2003da}.

In warped compactifications  $\mu_F$ can take any value below the 10D string tension $T_F$, 
but in brane-antibrane inflation models the tension is related to the inflation scale, 
with $G\mu \sim 10^{-6}$.  String networks surviving until today promise observable signals such as Cosmic 
Microwave Background (CMB) fluctuations and gravitational lensing.  Indeed, much  
excitement was generated by the observation of an unusual double galaxy, 
thought to be a candidate for
the first observation of a string lens \cite{Sazhin:2003cp,Sazhin:2004fv,Sazhin:2005fd},
but unfortunately ruled out by subsequent high-resolution observations  \cite{Sazhin:2006fe}.

The dynamics of networks even of ordinary cosmic strings in are not well understood. In the 
conventional scenario \cite{Hindmarsh:1994re,VilShe94} energy is lost from the network into 
loops, which oscillate and slowly decay into gravitational radiation.  However, when 
an underlying (classical) field theory with string solutions such as the Abelian Higgs model 
is simulated the energy goes into gauge and Higgs radiation via some kind of 
non-perturbative process acting at the string width scale \cite{Vincent:1997cx,Moore:2001px}. 
Whatever the energy loss mechanism, there is broad agreement that 
a network of cosmic strings formed in the early Universe will settle into a 
self-similar or so-called  \emph{scaling} regime, in which the gross length scales such as the 
average inter-string distance and the average curvature radius increase in proportion to time.

Scaling is an extremely important property, for it ensures that the strings contribute a constant 
fraction of the energy density and therefore do not come to dominate the Universe (or 
disappear to a negligible density).  The string density $\rho_s$ is approximately 
given by the string tension multiplied by the length per unit volume, neglecting 
kinetic energy.  On dimensional grounds, we can define a length scale $\xi$ such 
that $\rho_s = \mu/\xi^2$, and scaling means that $\xi \sim t$ or $\rho \sim \mu/t^2 $.  Hence 
dividing by the critical density $\rho_c \propto 1/Gt^2$ of a Friedmann universe, 
we see that $\Om_s \sim G\mu$.

However, once junctions are added to the network, it is not clear that scaling 
will be maintained, which would be cosmologically disastrous for the new cosmic string 
scenarios.  A ``frozen'' string network would act like a fluid with equation of state 
$p = -\rho/3$ \cite{Vilenkin:1984rt,Bucher:1998mh} which cannot be reconciled 
with current cosmological data. It is therefore extremely important to investigate the 
dynamics of such networks.

The first simulations of string networks with junctions 
\cite{Vachaspati:1986cc,McGraw:1997nx}
used very simple models through which it was hoped the dynamics of the field 
theory could be captured, and gave respectively negative and positive indications of scaling.
More recent attempts at modelling of string networks with junctions \cite{Tye:2005fn} give 
indications that scaling is allowed. 
Where there is an underlying field theory, it is to be expected that direct simulations of 
the field equations will give more reliable results.  Here again results are mixed. 
There is evidence from numerical simulations of 
domain wall networks in 2+1 dimensions both 
of departures from naive (linear) scaling \cite{Ryden:1989vj,Antunes:2003be} 
and consistency with it \cite{Avelino:2006xy}. 
The issue has been addressed recently in 3+1 dimensions
\cite{Copeland:2005cy}, reviving earlier work of \cite{Spergel:1996ai}, where
the non-linear sigma model (\nlsm) approximation of a field theory containing multiple vortices and
junctions was shown to scale with the background fluid. The
\nlsm\ approximation greatly reduces the number of dynamical variables and 
makes it possible to perform large-scale simulations. However, the model is somewhat peculiar 
in that the $\si$-model space is not in fact a manifold, and in any case misses the microphysics associated with the strings. 

Here we perform for the first time 
full field theory simulations for a model of a string network with junctions, specifically 
$\ZM_3$ vortices, and show that the
evolution is consistent with a late-time scaling regime. While our model does not 
fully capture the dynamics of $(p,q)$-strings, we can conclude that the presence of 
junctions is not in itself inconsistent with scaling and, by extension, that $(p,q)$-string networks are 
viable for sufficiently low $\mu_F$.

\section{The model}
\label{sec:model}
A field theory with spontaneous symmetry breaking from Lagrangian symmetry group $G$ to vacuum 
symmetry group $H$ 
has string solutions if the manifold $\M \simeq G/H$ (which may not coincide with the 
manifold of minima of the tree-level potential \cite{Hindmarsh:1992yy})
is not simply connected, or its first homotopy group is non-trivial: $\pi_1(\M) \ne 0$.  If $G$ is 
simply connected this can be ensured by arranging for $H$ to be disconnected: $\pi_0(H) \ne 0$. 

Disregarding the Abelian symmetry-breaking $G$=U(1), $H$=1, 
the smallest simply-connected compact Lie group is SU(2), which suggests its use as the 
symmetry of the Lagrangian of a field theory.  The most trivial discrete group is the cyclic 
group $\ZM_N$ and so we are led to the conclusion that the simplest model with string
junctions in a non-Abelian field theory would have a vacuum manifold SU(2)/$\ZM_N$.
To obtain a vacuum manifold SU(2)/$\ZM_N$ we use an adjoint scalar, $\Phi=\Phi^a T^a$ and a complex, 
$(2j+1)$-component vector $\phi$, where $j$ is the spin of the representation, which transform as
\ba
\Phi&\rightarrow& U\Phi U^{-1}\\
\phi&\rightarrow& U\phi
\ea

The action we use is given by
\ba
\label{eqn:action}
S&=&\int {\rm d}^4x ({\cal T}-\V)
\ea
where the kinetic energy density takes the standard form
\ba
{\cal T}&=&-\del_\mu\phi^\dagger\pa^\mu\phi-{\rm Tr}(\del_\mu\Phi\del^\mu\Phi)
\ea
and the potential is broken down into three pieces $\V ={\cal V}_1+ {\cal V}_2 + \V_3$, with
\ba
{\cal V}_1&=& \lambda_1({\rm Tr}(\Phi^2) - \eta^2)^2
            +\half\lambda_2(\phi^\dagger\phi - v^2)^2, \label{e:V1}\\
{\cal V}_2&=&-\lambda_3\eta\phi^\dagger\Phi\phi-\lambda_4(\phi^T C^T\phi+\phi^\dagger C^*\phi^\star)
             -\lambda_5(\phi^T C^T\Phi\phi+\phi^\dagger \Phi C^*\phi^\star),\\
{\cal V}_3&=& \la_6 |\phi^T C^T \phi|^2 + \la_7 \phi^\dagger\phi\Tr(\Phi^2),
\ea
where $C$ is the charge conjugation matrix (see Appendix \ref{AppA}).
The first potential, ${\cal V}_1$ is used 
such that the fields aquire a non-zero vacuum expectation value (vev). The second
potential is present in order to get the two fields to interact, and explicitly break
the residual U(1) symmetry of ${\cal V}_1$. The third part is included for 
completeness, and is not necessary to obtain the desired symmetry-breaking.  We accordingly 
set $\la_6$ and $\la_7$ to zero in the following. 
Note that if $j$ is half-integer, the charge 
conjugation matrix is skew-symmetric and so the $\la_4$ and $\la_6$ terms vanish.

\section{Vortex solutions}
\label{sec:solutions}
Our $(2j+1)$-dimensional SU(2) basis $T^a$ is chosen so that 
\ben
\Tr(T^aT^b) = \half \de^{ab}, \quad T^3 = \frac{1}{N_j} \mathrm{diag} ( j,j-1,\ldots,-j+1,-j),
\een
with $N_j^2 = \frac{2}{3}(2j+1)j(j+1)$ (see Appendix \ref{AppA}).

We can understand the presence of vortices in the model by considering the vacuum structure.
When the fields are in the vacuum, 
we can use two of the SU(2) rotations to get 
$\Phi=  h \eta T^3$, where $h$ is a dimensionless constant. 
Taking $\phi= v \left(a_{(j)},a_{(j-1)},...,a_{(-j+1)},a_{(-j)}\right)$
we find that the first term in ${\cal V}_2$,
\ba
-\lambda_3\eta\phi^\dagger\Phi\phi&=&- \frac{1}{N_j}\lambda_3 g \eta v^2 (j|a_{(j)}|^2+(j-1)|a_{(j-1)}|^2+...)
\ea
is minimized for a given $\phi^\dag\phi$ when $a_j$ is the only non-vanishing component. 
We can use the final SU(2) rotation to ensure that
$a_j$ is real giving the following ansatz for the vacuum
\ba
\Phi_0&=&h \eta T^3\\
\phi_0^T&=&v (a_{(j)},0,0,...)
\ea
Having established the form of the vacuum, we are in a position to find its residual symmetry.
Performing a rotation with $\tilde T^3=N_j T^3$, $U_\theta=\exp(i\theta \tilde T^3)$ we find
\ba
\Phi_0&\rightarrow&\Phi_0\\
\phi_0&\rightarrow&\exp(ij\theta)\phi_0
\ea
so we see that the vacuum is unchanged for $\theta=2\pi n/j$, which gives the 
discrete structure $\ZM_N$ we were aiming for, with $N=2j$. As we have a $\ZM_N$ structure for the
vortices then we see that there must be junctions. To understand this we
consider the case for $\ZM_3$, which is the model we shall actually be simulating.
In two space dimensions, a finite energy field configurations representing three vortices together 
can be continuously deformed to the vortex-free vacuum, while maintaining finite energy. Similarly, 
two anti-vortices can be continuously deformed to a vortex. If 
we make the deformation parameter to depend on a third space dimension, and 
we realised that there will be junctions where the transition takes place, presumably localised in 
order to reduce the total energy.  
One can also view the vortices as the bound-state of the anti-vortices of the other two, 
as for $(p,q)$ strings. 

Another way to understand junctions in  our model is by noting that we have an adjoint SU(2) scalar
field, and therefore when $v^2 < 0$ in the potential (\ref{e:V1}) the unbroken symmetry expands 
to U(1), and there will be monopole solutions \cite{'tHooft:1974qc}. An expectation value for the 
field $\phi$ breaks this U(1) down to $\ZM_N$, and confines the monopole flux to $N$ strings. 
Hence the monopoles become points where three vortices come together to form a junction. 
Note that in the gauged $\ZM_2$ case the monopoles are ``beads'' on strings \cite{Hindmarsh:1985xc}.

\FIGURE[ht]{
\epsfig{file=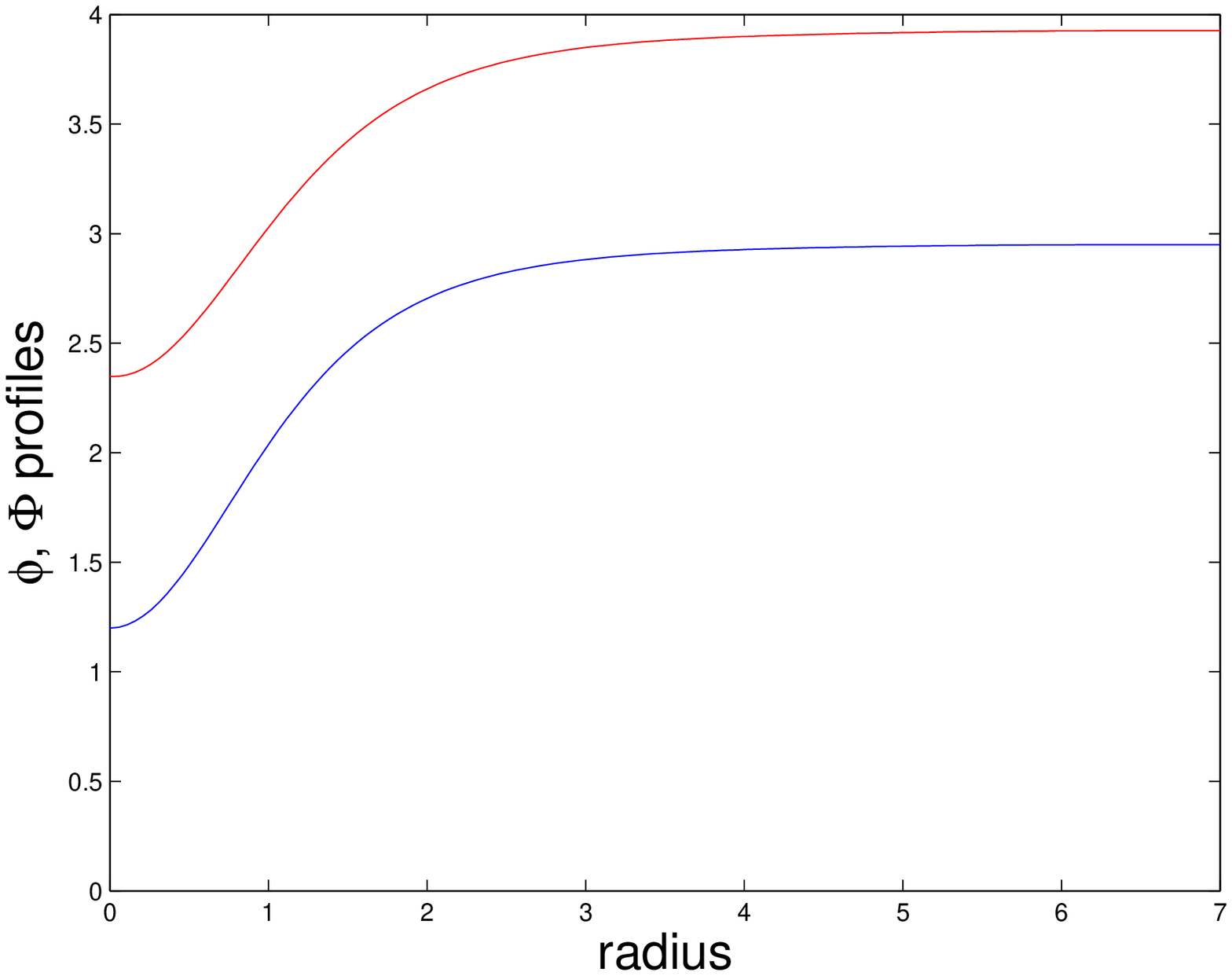,width=10cm}
\caption{
A representation of the vortex profile functions $|\phi|^2$ and $\Tr\Phi^2$, 
with parameters given in Eq.\ (\ref{eqn:parameters})
\label{fig:profiles}
}
}

This is far from being the only way to obtain vortices with junctions from a field theory. Indeed, 
the Abelian symmetry breaking U(1)$\to$1 has a spectrum of vortices labelled by the integers, 
and stable bound states of vortices exist both in the pure scalar theory, and in the gauge theory when the gauge boson is lighter than the Higgs boson (corresponding to a Type II superconductor). Recent work has shown that string networks with junctions can form in this theory \cite{Donaire:2005qm}, but not whether the networks scale.   

A model closer to the $(p,q)$-string, which has two separate U(1) symmetries corresponding to 
the electric flux of the F-string and the magnetic flux of the D-string, has recently be proposed 
by one of the authors \cite{Saffin:2005cs}.  There are two U(1) symmetries in this model, but 
both kinds of string have magnetic fluxes, and the characteristic string tension formula 
(\ref{e:PQtension}) is not exactly obeyed. 

A more complex model based on a gauged U(N) with a global SU(N) symmetry has also been put forward \cite{Hashimoto:2005hi}.  The global SU(N) symmetry can be lifted in such a way as to give 
string junctions, and the strings have the very interesting feature that it is possible for different 
kinds to pass through one another without reconnecting.

\section{Evolution}
\label{sec:evolution}
Ideally we would like to perform the simulations in the background of some standard
cosmology, typically a radiation or matter dominated Friedmann model. 
In practice it turns out that for
field theories with an explicit mass scale this is problematic. In order to properly simulate the vortices
one must choose a sufficiently small lattice spacing to resolve their core throughout
the whole simulation. In an expanding background, the physical separation between lattice
sites increases with time, so the vortices effectively ``fall through the lattice''
after some period of time. The other effect of expansion is Hubble damping, which is
easy to simulate. In our simulations we wish to mimic the effect of expansion in a radiation
dominated Universe by first transforming to conformal time, $\tau$, the equations of motion
for a real scalar with potential $\V$ then become
\ba
\varphi''+2\frac{a'}{a}\varphi'-\nabla^2\varphi&=&-a^{2\alpha}\frac{\partial}{\partial\varphi}{\cal V},
\ea
where the prime denotes differentiation with respect to $\tau$, $a(\tau)$ is the scale factor, $\nabla^2$ is the co-moving Laplacian, and $\al$ is a constant which should be 1. However, as explained above this causes problems at late time with the vortices not being
resolved on the lattice. Instead we adopt a tactic suggested in \cite{Press:1989yh}, 
which is to reduce the rate at which the strings shrink by reducing the size of the parameter 
$\al$.  In our simulations we take $\al = 0$, which results in vortices of constant comoving size,
we also evolve with the Hubble damping appropriate to a radiation dominated Universe.
According to \cite{Press:1989yh}, this approximation has little effect on the dynamics of the 
vortices themselves, at least in the U(1) model they considered.

String networks are thought to quickly evolve towards a statistically simple 
self-similar scaling solution, where
all the relevant length scales in the problem evolve proportional to time. 
What we aim to determine here is whether non-Abelian networks of the type
constructed above follow the scaling rule. The
most elementary scale of the problem is the average separation of strings, $\xi$,
and this can be constructed by knowing the average comoving length $L$ of string per unit
co-moving volume $V$ as
\ba
1/\xi_s^2&=&L/V.
\label{e:XiDef}
\ea
Scaling would mean that $\xi_s\propto\tau$. 

Determining the length of string in this model in a lattice simulation 
is not as simple as for the Abelian model with a 
simple complex scalar field, where the field vanishes in the core of the string in the 
continuum.  
In that case we can identify plaquettes through which a string passes by use of the 
geodesic rule, whereby the phase between two neighbouring lattice points is assumed to interpolate 
between the two values using the shortest path.
No such convenient local indicator of the string position is guaranteed to exist on a lattice, 
so we instead use the profile functions of the vortices, Fig. \ref{fig:profiles}, noting that 
$|\phi|$ is at a minimum (but not zero) at the centre. We therefore identify the length of 
a string by the volume of the region where $|\phi|^2<|\phi_c|^2$, divided by the 
area of the static solution with $|\phi|^2<|\phi_c|^2$, 
for some cutoff $|\phi_c|^2$. In the early stages of the
simulation we would not expect this to be a good measure due to the large oscillations of the
field, however, as the simulations proceeds these oscillations decay away thanks to Hubble damping, leaving a cleaner
signal. In practice we took three different cutoffs, finding that there was no significant
difference between them, lending credence to this as a reliable measure of string length.

For the junctions we can construct a similar length scale using dimensional analysis,
\ba
\label{eqn:xim}
1/\xi_m^3&=&N/V,
\ea
and this corresponds to the typical distance between junctions.
The subscript $m$ is used because the junction charge corresponds to the monopole charge
of the adjoint scalar $\Phi$, $N/V$ is then the number density of monopoles.
Unlike the string charge, we can evaluate the integer monopole charge on a lattice, and
in the simulations we use this to measure $N$.

We now come to the details of the simulations.
The equations were solved using a simple 3-point discretisation of the Laplacian and 
a leapfrog time evolution, with the code parallelised using the {\sc lat}field 
library for lattice field evolution \cite{latfield-web-page}.
We used a $544^3$ box with the scalar potential parameters
\ba
\label{eqn:parameters}
\lambda_1=3/4,\quad\lambda_2=1,\quad\lambda_3=N_j/j,\quad\eta=\nu=1
\ea
and the rest either irrelevant or vanishing, as discussed above. The lattice spacing was taken as
$h_x=0.4$ and the time step given by $h_t=0.3h_x$. The simulations were run up until the time
it takes for light to cross half of the box, using a radiation background as discussed earlier.
For initial conditions we used the field configuration which resulted from setting the fields to their
vacuum value with random orientations, then damped to remove the large amount of radiation.
The damping lasted for 200 time steps and the damping factor during this period was $\Gamma=2.0$,
where the damping is included as $\phi''+\Gamma\phi'+...$;
the half-box crossing time includes this damping period.

\FIGURE[bt]
{\epsfig{file=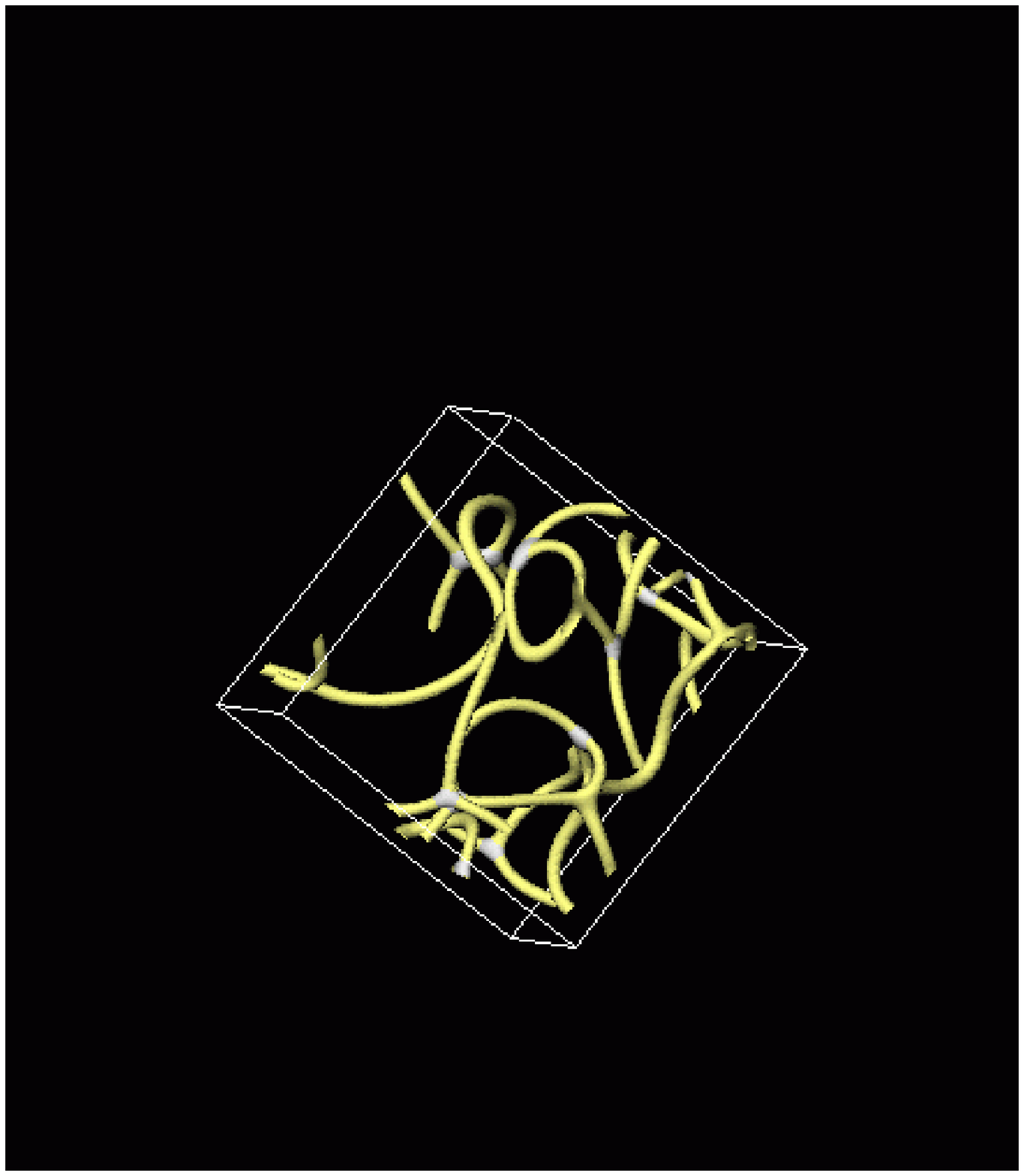,viewport = 70 80 370 350, clip}
\caption{
a snapshot of a run, showing the three-point vertices in a run using the parameters
of (\ref{eqn:parameters}).
\label{fig:snapshot}
}
}

We performed nine realizations with
a typical snapshot as shown in Fig.\ \ref{fig:snapshot}, where we can clearly make out the
prescence of vortices and junctions.
The result of these simulations is given in Fig.\ \ref{fig:scaling}, showing the mean evolution
of the length scales $\xi_s$ and $\xi_m$, with 1-$\sigma$ statistical errors either side. 
From this plot we see the linear
behaviour of both length scales indicating scaling behaviour. To evaluate the gradients, $\dot\xi$, we
present Fig.\ \ref{fig:scaling} showing that at late times the gradient is consistent
with being a constant, implying scaling for $\xi_s$ and $\xi_m$,
with scaling value extracted from the last half 
of the simulation 
\ba
\xi_s &=& (0.4\pm 0.1) \tau,\\
\xi_m &=& (0.5\pm 0.2) \tau.
\label{e:XiScale}
\ea

\FIGURE[bt]
{\epsfig{file=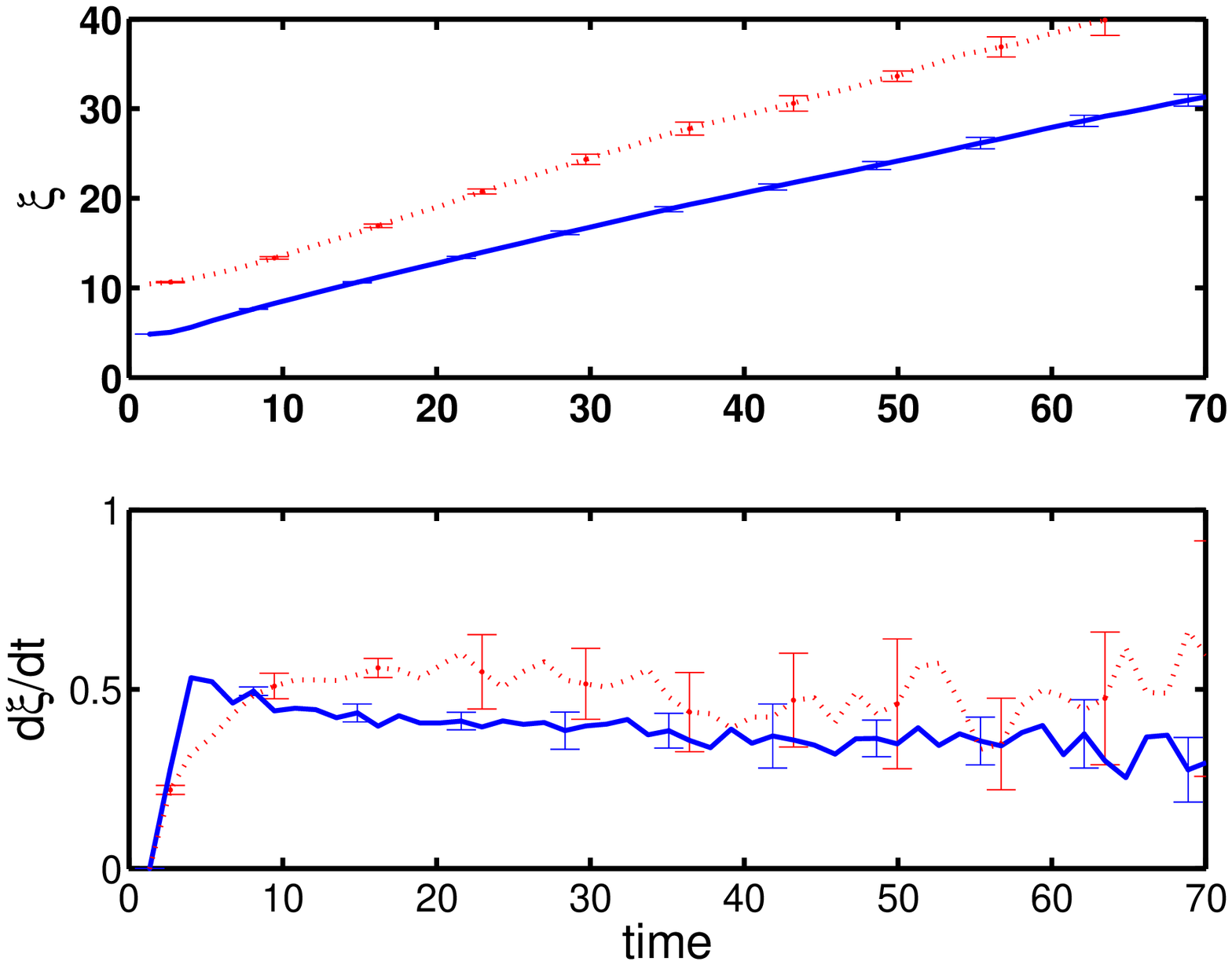,width=0.8\textwidth}
\caption{Top: network length scales $\xi_s$ (solid) and $\xi_m$ (dashed), against conformal time $\tau$, averaged over nine realisations. Bottom: the time derivatives of the network length scales, again averaged over nine realisations.
\label{fig:scaling}
}
}

\section{Conclusion}
\label{sec:conclusions}

We have carried out numerical simulations of a classical field theory containing strings with 
junctions, in order to test whether the presence of junctions in the network affects scaling.  
Our model is based on the global symmetry breaking SU(2)$\to$U(1)$\to\ZM_N$, and the junctions 
can be thought of as global monopoles connected by global strings.
We find good evidence, displayed in Fig.\ \ref{fig:scaling}, that the string network 
length scale $\xi_s$ (defined in Eq.\ \ref{e:XiDef}) scales in proportion to time with coefficient $0.4\pm0.1$
and the junction length scale $\xi_m$ (defined in (\ref{eqn:xim})) scales with coefficient $0.5\pm0.2$.

While our model was motivated by an attempt to model $(p,q)$ string networks, it fails to capture 
some important features:the spectrum of string tensions differs, with the field theory
giving equal tension to all vortices; the
reconnection probability is close to or equal to unity.
Our conclusions are that it 
is perfectly possible for networks with junctions to scale, 
backing up the results for the NL$\sigma$M
approximation in \cite{Copeland:2005cy},
and that they do not in themselves 
constitute a cosmological disaster. However, we must wait for better models before drawing 
quantitative conclusions for $(p,q)$ string networks.

\acknowledgments
We gratefully acknowledge the extensive use of the UK National
Cosmology Supercomputer funded by PPARC, HEFCE and Silicon Graphics.
PMS is supported by PPARC.

\appendix

\section{SU(2) conventions}
\label{AppA}
To construct the spin-$j$ representation of SU(2) we begin with a 
basis for the algebra $J_\pm$, $J_3$ normalised in the usual way
\ba
\lbrack J_3,J_\pm\rbrack&=&\pm J_\pm,\\
\lbrack J_+,J_-\rbrack&=&2J_3.
\ea
We identify the Casimir $J^2=J_- J_+ +J_3(J_3+1)$ and construct 
normalized eigenvectors
of $J^2$ and $J_3$,
\ba
J_3|\beta,m\>&=&m|\beta,m\>\\
J^2|\beta,m\>&=&\beta|\beta,m\>\\
\<\beta,m|\beta,m\>&=&1
\ea
By considering the highest and lowest weights one finds the following matrix
components for the generators $\tilde{T}^1$
\ba
\tilde{T}^+_{m,\tilde{m}}&=&\<j,m|J_+|j,\tilde{m}\>=\sqrt{(j-\tilde{m})(j+\tilde{m}+1)}\delta_{m,\tilde{m}}\\
\tilde{T}^-_{m,\tilde{m}}&=&\<j,m|J_-|j,\tilde{m}\>=\sqrt{(j+\tilde{m})(j-\tilde{m}+1)}\delta_{m,\tilde{m}}\\
\tilde{T}^3_{m,\tilde{m}}&=&\<j,m|J_3|j,\tilde{m}\>=\tilde{m}\delta_{m,\tilde{m}}
\ea
with $\beta^2=j(j+1)$ and $-j <m,\tilde m<j$.
Defining $\tilde T^1 = (\tilde T^+ + \tilde T^-)/2$, $\tilde T^2 = (\tilde T^+ - \tilde T^-)/2i$ 
we have
\ba
\sum_a \tilde T^a \tilde T^a&=&j(j+1). 
\ea
In our model we rescale by $N_j$, where
\ba
N_j^2=\frac{2}{3}(2j+1)j(j+1),
\ea
to get
\ba
T^a&=&\frac{1}{N_j}\tilde T^a,\\
\Tr(T^aT^b)&=&\half\delta^{ab}.
\ea
In the model we also use the charge conjugation matrix $C$ which is the intertwiner
between the SU(2) representations $T^a$ and $-T^{a\star}$,
\ba
CT^aC^{-1}&=&-T^{a\star}
\ea
So, by defining SU(2) group elements as $U=\exp(i\theta^aT^a)$ we have
\ba
CU=U^\star C.
\ea
In our basis we take
\ba
C&=&\left( 
      \begin{array}{cccc}
        \ldots & 0 & 0  & 1 \\
        \ldots & 0 & -1 & 0 \\
        \ldots & 1 & 0  & 0 \\
               & \vdots & \vdots  & \vdots
      \end{array}
    \right)
\ea

\bibliographystyle{JHEP}

\bibliography{c_string}

\end{document}